# Implementation of an Energy Management System for Real-Time Power Flow Control in AC Microgrid

Airin Rahman, Hafte Hayelom Adhena, Ramy Georgious, Pablo Garcia

Dept. of Electrical Engineering

University of Oviedo

Gijon, 33204, Spain

Email: airin.r82@gmail.com, hafteh24@gmail.com, georgiousramy@uniovi.es, garciafpablo@uniovi.es

*Abstract*— **Microgrid (MG) system, which is composed of renewable resources with the utility grid, energy storage unit, electric vehicles, and loads, acts as a single controllable entity. To get efficient and low-cost energy, need to manage power flow within MG depending on renewable resources and load demand. This paper proposes an energy management system (EMS) for grid-connected photovoltaic (PV) and energy storage system (ESS) based on AC MG. The proposed EMS regulates power flow through the AC bus considering next-day solar irradiance prediction, tariff profile, and day-ahead load demand to minimize the cost of electricity. The recommended EMS for the MG system has been verified using MATLAB/Simulink. Also, the algorithm has been tasted with real time data of a 10kW PV system.**

## I. Introduction

The increasing electricity demand due to the large number of populations, limited conventional resources, and concern about environmental issues has prioritized the use of renewable energy sources. However, the unpredictable nature of renewable sources has come with the concept of a microgrid (MG) system, which is a hybrid combination of distributed generation (DG), grid supply, electric vehicles (EV), and energy storage systems (ESS). Power electronic devices are playing a crucial role to make the MG system smarter and more flexible by managing the weakness and stability issues [1]. To meet the dynamic response and high reference tracking characteristic of those power electronic devices, control methods are vital concerns [2]. In the literature, several control schemes have been found which are ensuring the control of voltage and frequency along with the power-sharing in AC distribution networks [3], [4], and DC grids [5], [6]. Numerous types of control approaches have been done to overcome the limitation and improve the reliability of the MG system.

The energy management system (EMS) is considered one of the most promising options to enhance the efficiency of the microgrid system. Nowadays EMS is demanding significant research interests to make sure the optimal use of energy and reduce cost. Among several EMS, most of them are either focusing on energy storage lifetime or controlling of load to reduce expenses. The DC bus source connection topology is maintaining synchronization issues to extend battery life [7]. Optimal use of PV array production for the reduction of fast power charges in the battery is decreasing the electricity bills [8]. The optimal energy management system has also been designed to minimize energy cost considering the use of electricity with tariff time [9]. The fuzzy logic control system has been used to save energy by maximizing the power supply from solar and battery [10]. Another popular EMS concept is focusing on residential side power management, a home energy saving system is using for maintaining monthly electricity bills [11] and a smart home energy management system is facilitating the load management at the supplier side [12]. An intelligent algorithm has been developed to extract maximum solar power as an alternative to conventional solar charge controllers [13]. Advanced EMS is considering both forecasting and optimization to schedule day-ahead power generation and load demand in [14], here the EMS is operation through the optimization of load utilization based on the forecasting. But optimization of load demand is not a good solution for all cases, specially for EV connected MG system, where the connecting EV is expecting to consume power. This paper deals with the solution of this type of MG system through avoiding load optimization.

Most of the EMS is focusing on load management and energy storage to improve efficiency and reduce costing. Besides that, selling additional DG power to the grid can also be a way to reduce costs. This work is proposing an algorithm for the MG system to reduce expenses by selling additional power generation and scheduling battery charging period to use either DG power or grid power during lower tariff period. The EMS will consider solar irradiance forecasting, tariff, and load demand profile to decide the system operation stage. The algorithm will maintain the power flow of the AC microgrid to minimize the electricity bills by controlling the charging period of the energy storage system as well as by providing the additional PV generation to the grid. Here the forecasting for both solar irradiance and load demand is considering to manage the power flow. The suggested EMS is verified using the MATLAB Simulink model, where the simulation has done to deal with the real-time data to verify the effectiveness of the proposed algorithm. Section II introduces the grid-tied MG. Section III explains the proposed EMS algorithm. Section IV presents the simulation model and section V represents the simulation result. Section VI contains the simulation result with real-time data. Finally, section VII states the conclusions and future work.

## II. Grid-Tied System for Energy Management

The MG system in this work is operating on a grid-tied model, which is based on AC side controlling. PV array is considered as the DG, Li-ion battery arrangement is being used as the energy storage system (ESS), where some of the batteries could be batteries of electric vehicles. Power electronic converters are utilized for the interconnection of PV and ESS with the grid and loads. The ESS is connected through a bidirectional inverter. The EMS will control the power flow in the AC bus so that the PV system can supply power to both load and grid when the PV generation is higher than load demand, also the power flow for battery charging will be controlled based on the solar irradiance next-day forecasting. The simulation has been done with PV array module and battery as ESS, where the power range of PV array is 20KW to 40KW and the stage of charge (SoC) range of the battery is between 10 to 90 percentage. The power rating of the inverters is 5kW. Figure 1 shows the basic layout of the grid-tied MG system which has been used to verify the proposed EMS.

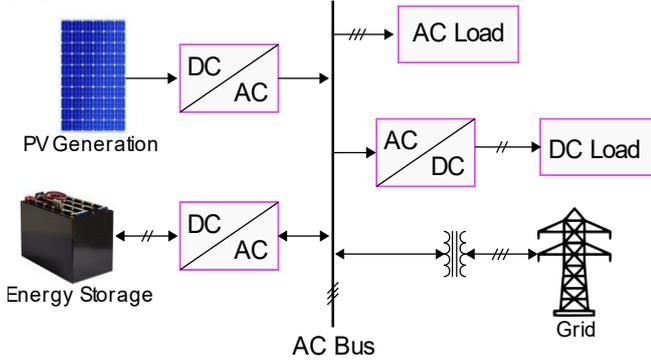

Figure 1. Grid-tied AC microgrid system.

### III. PROPOSED ENERGY MANAGEMENT SYSTEM

The proposed energy management system is controlling power flow through the AC bus to decide the charging period of the ESS considering solar irradiance forecasting and the use of additional PV generation so that the extra power generated by DG can also provide to the grid. The EMS will help to reduce the electricity bills by charging the ESS during a cheap tariff period or from the PV generation and also by selling the additional PV generation to the grid.

In the EMS, EV is considering as load. So, the first step of the EMS is to check the status of EV and calculate the total load power. In the beginning of the algorithm, the system will check whether EV is connected to the MG or not at that time. If the EV is connected, then the system will check the SoC of EV's battery and if the value of SoC suggests that the EV's battery need to charge, total load power will be the summation of load demand and required power of the EV. Equation (1) represents the total power calculation.

$$P_{L\_total}(t) = P_l(t) - P_{EV}(t) \quad (1)$$

where $P_{L\_total}$ is the total load power, $P_l$ is the load demand, and $P_{EV}$ is the electric vehicle power.

If the EV is not connected or the SoC of EV's battery is greater the maximum SoC limit of the battery, then the total load power will be equal to the load demand. The calculated load power is used to take decision in the proposed algorithm through comparing with the PV generation.

The proposed algorithm has four cases depending on PV generation, total load power, and tariff profile to operate at six different modes of operation by controlling the power flow. The algorithm is taking PV power generation, load demand, EV power, and tariff profile as the input data to make the decision about the operating modes considering the battery SoC level, feed-in tariff (FiT) value, and next day solar irradiance (G).

Figure 2 shows the flowchart of the proposed algorithm, here 'G' is next day solar irradiance, 't' is the time variable, and 'm' is the step size variable.

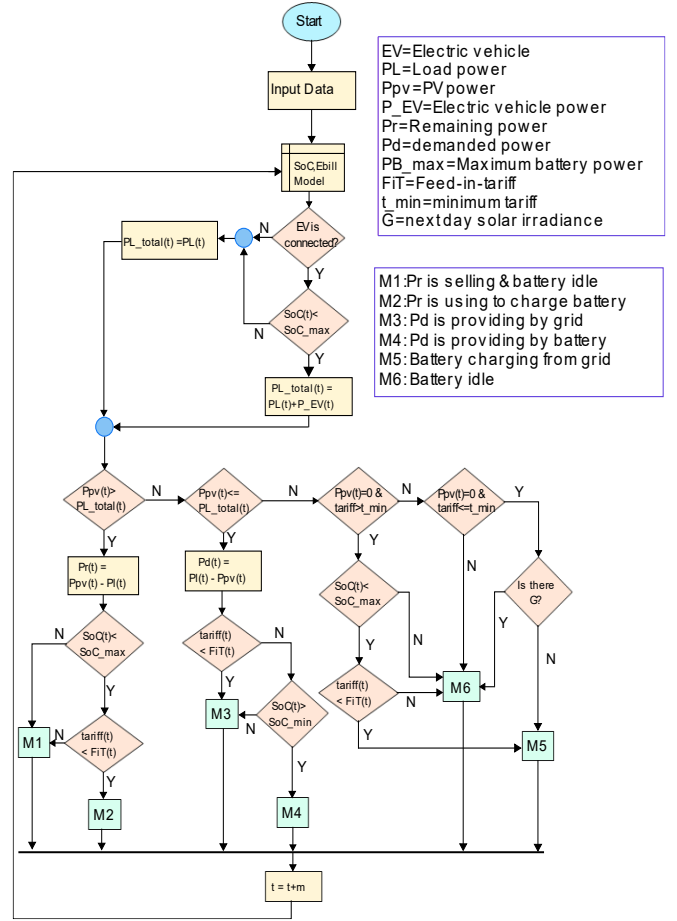

Figure 2. Flowchart of the proposed algorithm.

The first case is considered when the PV generation will be greater than the total load power. So, all the consumption will be provided by PV generation and the remaining power is calculated by equation (1).

$$P_r(t) = P_{pv}(t) - P_{L\_total}(t) \quad (2)$$

where $P_r$ is the remained power, $P_{pv}$ is the solar power generation, and $P_{L\_total}$ is the total load power.

In this case, the system will check the SoC of the ESS (Li-ion battery) to decide whether the charging of the battery is needed or not. Also, EMS will check the tariff before deciding to charge the battery. The battery will only charge if the tariff of that moment is less than the FiT, otherwise, the remaining power will sell. Mode 1 (M1) will keep the battery in idle condition and provide all the additional power to the grid. Mode 2 (M2) will consider as a battery charging period, so the remaining power will use for battery charging. If the SoC of the battery is lower than the maximum level but the tariff of that point is not lower than FiT, in that case, the system will stay on M1 and the remaining power will feed to the grid to take the benefits of high tariff value by selling DG power.

The second case is when the PV generation will be lower than the total load power. In this case, the PV generation is not enough to supply the whole consumption, so the additional required power is calculated through equation (2).

$$P_d(t) = P_{L\_total}(t) - P_{pv}(t) \qquad (3)$$

where $P_d$ is the demanding power. This case will decide either the ESS or utility grid will supply the power needed to meet the entire load power. Mode 3 (M3) is the period of operation when the grid will entirely supply the demanded power calculated in equation (2). The system will operate at this mode if the tariff of that time is lower than the FiT value. The system will enter mode 4 (M4) if the tariff is not lower than the FiT and the battery SoC is higher than its minimum level, in this mode demanded power will be provided by ESS.

The third case is considering the situation when there will be no PV generation, but the tariff is not the lowest. In this case, the battery charging period will depend on the tariff value. The system will check the tariff with FiT value and if the tariff is lower than FiT, the system will enter mode 6 (M6) to charge the battery using the lower tariff value. Otherwise, the system will be on mode 5 (M5) by keeping the battery in idle condition.

The fourth case of the algorithm is the situation when there will be no PV generation and the tariff will be the lowest, mostly in the night-time condition. In that case, the system will check solar irradiance forecasting of the next day. If the next day solar power generation is more than enough for the load demand then the system will operate at mode 5 (M5), where the battery will be in idle condition and wait for charging on the next day. However, if the forecast shows that the next day solar irradiance will not be enough to generate the total expected load demand, in that situation the system will enter mode 6 (M6) to charge the battery. Where the battery will charge from the grid using the lowest tariff profile.

## IV. SIMULATION OF MICROGRID SYSTEM TO VERIFY THE PROPOSED EMS

The simulation of the grid-tied MG system was done for the verification of the proposed EMS algorithm in MATLAB/Simulink. As the controller will design for microgrids with distributed energy resources, the average model inverter has been used instead of the switching model for simplicity. Vector current controller has designed as the controller of the inverters connected to the PV and the battery. The EMS system control is focused on power flow managing through the AC bus, the active and reactive power values have been used to generate the reference signals for the controller. Where the reactive power is kept at zero value. Table 1 shows the parameters used in the simulation model.

TABLE I. PARAMETERS USED FOR SIMULATION MODEL

| System Parameters | Values |
|---|---|
| Grid Voltage (Van) | 400V |
| Frequency (f) | 50Hz |
| PV Power | 10kW |
| Load Power | (2-8) kW |
| EV Power | (0-5) kW |
| DC Battery Voltage | 650V |
| Battery SoC Range | 10 % – 90 % |
| Inverter Power | 5kW |
| Filter Inductance ($L_f$) | 7mH |
| Filter Resistance ($R_f$) | 0.1Ω |
| **Control Parameters** | **Values** |
| Damping ratio (ξ) | 0.707 |
| Natural frequency ($w_n$) | 2π300 rad/s |
| Kp, Ti | 18.6573, 1.33×10³ |

### A. Simulation with Average Model Inverter

The simulation of the MG system has done using the average model inverter to reduce the complexity of the simulation and make response faster, which also maintains sufficient dynamic accuracy by maintaining the same average V-I terminal relationship on AC and DC side. Based on the controller functionality average model is identical to the switching model, while the power stage shows the only key difference. In order to emulate inverter behavior, three single controlled voltage sources are used in the three-phase average model inverter instead of using three-phase voltage source inverters like switching mode. The average model offers less numerical convergence problem, as the state variable of the model is averaged over the switching period.

### B. Design of Current Controller for AC Microgrid

A standard control strategy, vector current controller has been designed for the controlling of the AC microgrid system. The design of a vector current controller is based on the synchronous rotating reference frame, which enables the PI controller to control the active and reactive powers indirectly. Since the synchronous rotating frame transforms the AC values to DC through coordinate transformation, active and reactive power control is performed through d-q axes current controlling. A phase-locked loop (PLL) is used to maintain the synchronization of d-q axes current with the grid voltage. PLL extracted the phase angle of the grid voltage and provides that as the phase angle for the Park transformation to generate the d-q axes current [15]. Figure 3 shows the schematic diagram of a standard current controller.

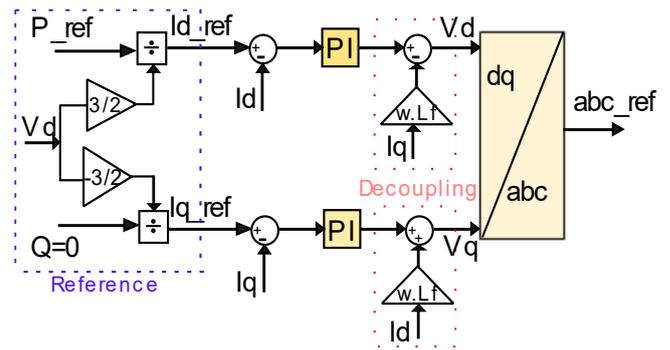

Figure 3. Vector control current controller for PV and battery inverters.

The controller has been designed to regulate the d-q axes current considering both feedforward and feedback. Tuning of the PI controller has been done by taking filter inductance as the plant of the system. The generated reference signal from the PI controller is used to calculate voltage and current reference value for the controlled voltage source and controlled current source respectively of the average model inverter. To get reference voltage for the controlled voltage source, the reference signal is multiplied by a DC voltage, and the controlled current source reference current is generated by multiplying with the AC side current to maintain the V-I relationship of the average model inverter.

The d-q axes currents reference signals have been generated by taking active and reactive power reference values. Since in the synchronous frame, d axes components are coincident with the instantaneous voltage vector and the q axes component is in quadrature with it, the reference part of Figure 3 shows the generation of d-q current reference signals [16]. Equations (3) and (4) define the active and reactive power relation in the d-q frame.

$$P = \frac{3}{2} V_d I_d \quad (3)$$

$$Q = -\frac{3}{2} V_d I_q \quad (4)$$

Two identical control systems have been designed for both inverters connected with the PV system and ESS.

## V. SIMULATION RESULTS

To verify the algorithm, in the first step the simulation has performed with some random values of PV generation, load demand, and EV power. The simulation has been performed with the data for complete one day by considering different situation to justify the modes shift of the EMS.

First, the simulation has been performed considering that the EV is connected with the MG system throughout all day with different power profile and the SoC of EV's battery is lower than the maximum level of it. So, the total load is calculated using equation (1) and that calculated total load value has been used to decide the case of the EMS.

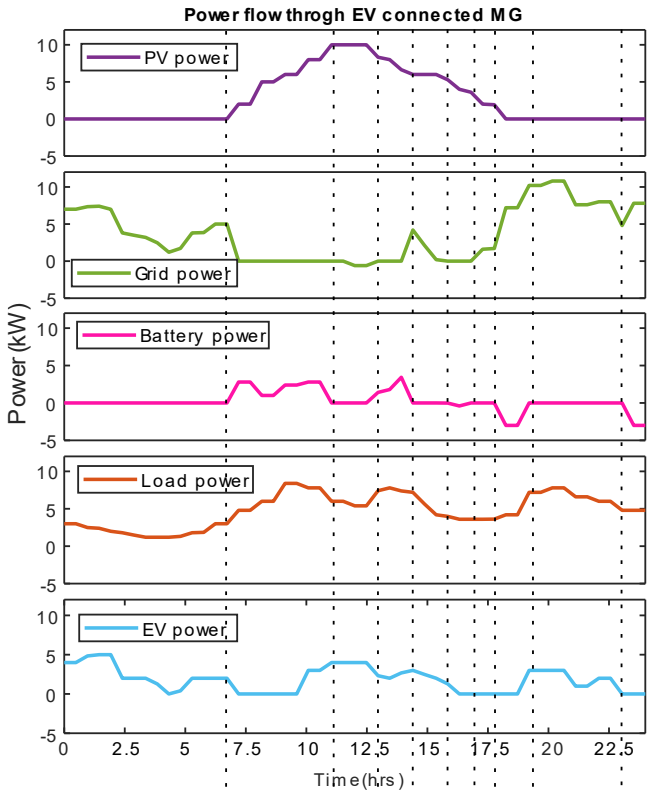

Figure 4. Power flow through the MG considering the EV is connected all day.

Figure 4 demonstrates the power flow the MG, which indicates the mode shifting based on the proposed algorithm and shows that the system is operating in all modes. When there is no PV generation, grid is providing total load power and depending on the next day solar radiation and tariff profile the system enters on M6, where the battery is staying on idle condition. Then the system is operating in M4, as the PV generation is lower than total load power and the tariff profile is not lower than FiT value, so the battery is discharging to supply total load power and no power is consuming from the grid in this mode; Figure 4 indicating that grid power is zero and the positive power flow through battery specifies discharging mode. When the PV generation goes higher than the total load power, depending on the proposed algorithm, the system enters on M1. The negative grid power and no power flow through battery indicates that addition PV power is selling to grid and the battery is in idle condition. Based on the proposed algorithm, M2 executed if the battery needs to charge, and the tariff is lower than FiT. In the time period of M2 of Figure 4, shows that the grid power is zero as the total load power is providing by PV generation and negative battery power reveals that the additional PV generation is using for battery charging. After that as the PV generation goes lower than total load power and the tariff is less than FiT, the system enters M3 where the demanding power to satisfy total load power is coming from the grid. Again, when the PV power is equal to zero, total load power is providing from grid and depending on the tariff profile the system will operate either on M5 or on M6. In M5 battery is charging from the grid and in M6 battery stays in idle condition.

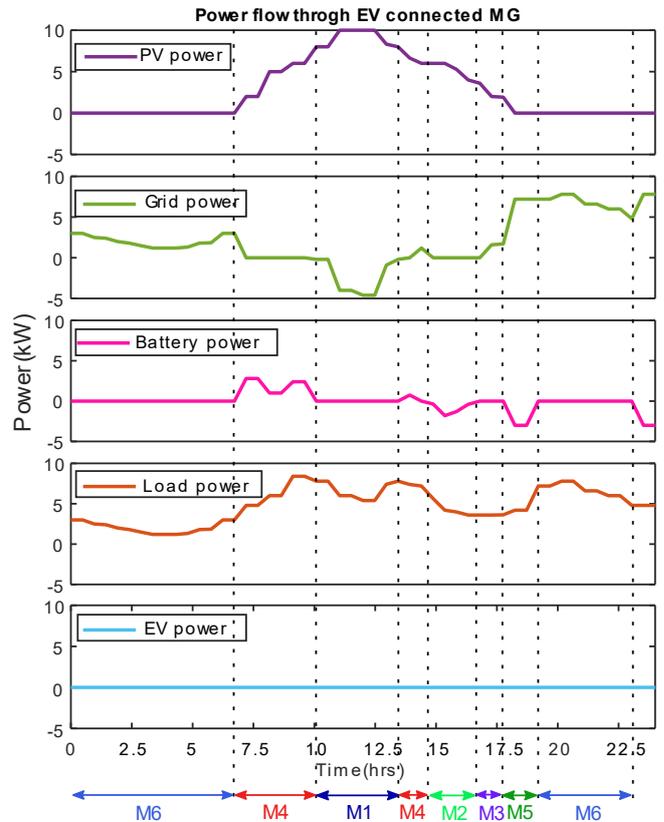

Figure 5. Power flow through the MG considering EV is not connected.

Again, the simulation has been performed with same tariff and SoC profile considering that EV is not connected throughout all day. So, in that stage, total load power is equal to the load demand. Figure 5 represents the power flow of the MG system, where EV power is equal to zero as there is no EV connected with the system throughout the day. As the tariff and SoC profile are same and the comparison between PV generation and load demand is not changing cases, so the system is shifting modes with previous sequence. Figure 5 shows the same mode shifting order like Figure 4, the only difference is the amount of power consuming or supplying to the grid and battery. Figure 5 shows that in M6 consumption of grid power is lower than the previous simulation result, as the electric vehicle power is zero. Also, the amount of selling power in M1 and battery charging power in M2 is significantly higher than the previous simulation result.

Also, the simulation has been performed considering that there is no need to charge the ESS throughout all day. As the SoC level of the battery is higher than the maximum SoC level of the battery, the system will not enter to M2 and M5 to charge the battery. Figure 6 shows the power flow through the MG and indicates the operating mode of the system.

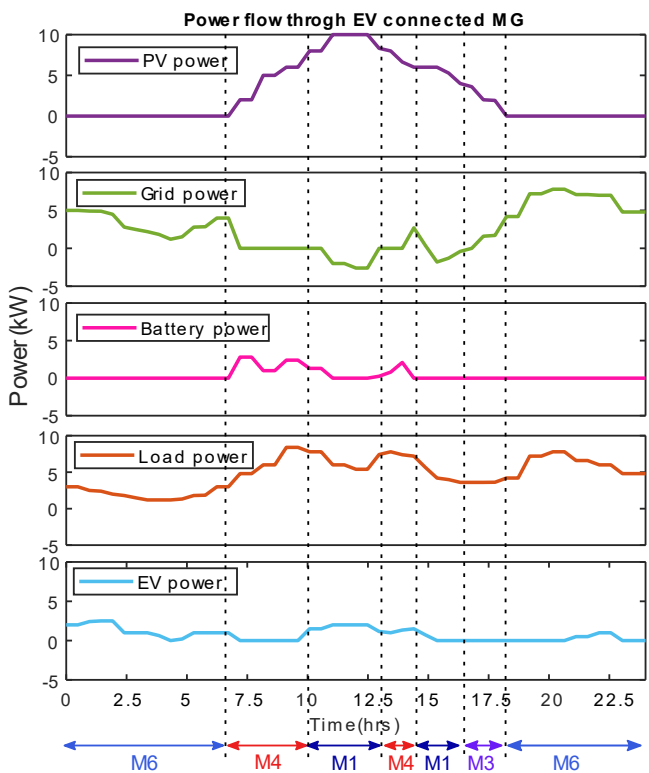

Figure 6. Power flow through the MG considering that no need to charge ESS.

Figure 6 represents that the battery is in inactive condition with zero power flow at M1, M3, and M6. The positive power flow of the battery in M4 indicates that the battery is discharging to provide total load power as the tariff profile is greater than FiT and battery has the capacity to supply. In M1, the negative power flow through the grid presents that the addition PV power is selling to the grid. Comparing with previous simulation, in this consideration the system enters to M1 two times. As the battery SoC is bigger than its maximum level, instead of using the additional PV power to charge the battery, the remaining power is selling to the grid.

Again, the simulation has been performed considering a day with bad weather condition by taking the PV power generation lower than previous simulations PV power, but keeping the load demand and EV power same. In this combination of power generation and load power, the system doesn't enter to case 1, as the PV generation never goes higher than total load power. With this combination of input data, the system remains on M4 for longer duration, where the battery is discharging to supply the required power as the PV generation is less than total load power. After that, when the tariff profile goes lower than FiT value, the system shifts to M3 where the power flow through the battery is zero and the demanding power is providing from grid. When the PV generation is equal to zero, the system stays a longer time in M5 as the tariff is lower than FiT and the next day solar radiation is suggesting that there will be not enough PV power generation to cover the load demand. The negative power flow through the battery represents battery chagrining period. As in M5 the battery is charging by taking power from the grid, the grid power at that M5 goes higher than 10kW. Also, when the tariff value goes higher than FiT, the system operates on M6. Figure 7 demonstrates the power flow through the MG considering bad weather condition, where the power flow through the grid and battery justify all the mode shifting.

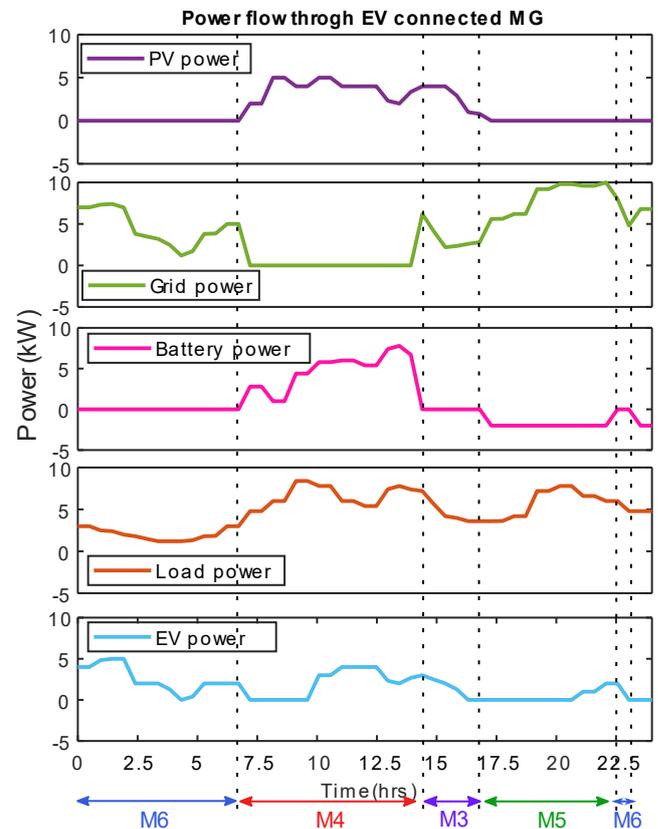

Figure 7. Power flow through MG considering bad weather condition.

## VI. IMPLEMENTATION OF EMS WITH REAL-TIME DATA

After completing the simulation with some random value which is maintaining the power flow according to the proposed algorithm, the simulation of the MG system has been performed with real-time data. Figure 8 represents the simulation result with real-time data. The simulation has been performed using one-day data, taking one hour as the time difference. Figure 8 shows that in M6, total load power is proving from the grid and when the PV generation goes higher than total load power the system enters into M1 to selling the additional PV generation as the tariff profile is higher than FiT. After that the system enters in case 2 and check the SoC and tariff profile to choose whether grid or battery will supply the required power to fulfill total load demand. As the tariff profile is higher than FiT at that during, in case two the system operates on M4 where the battery discharge to supply demanding power. When the tariff goes lower than FiT value, the system shifts to M3 to supply the required power from the grid. Since the next day solar forecasting indicates that the PV generation will be enough for the load demand and battery charging, the battery stays is in idle condition throughout the nighttime.

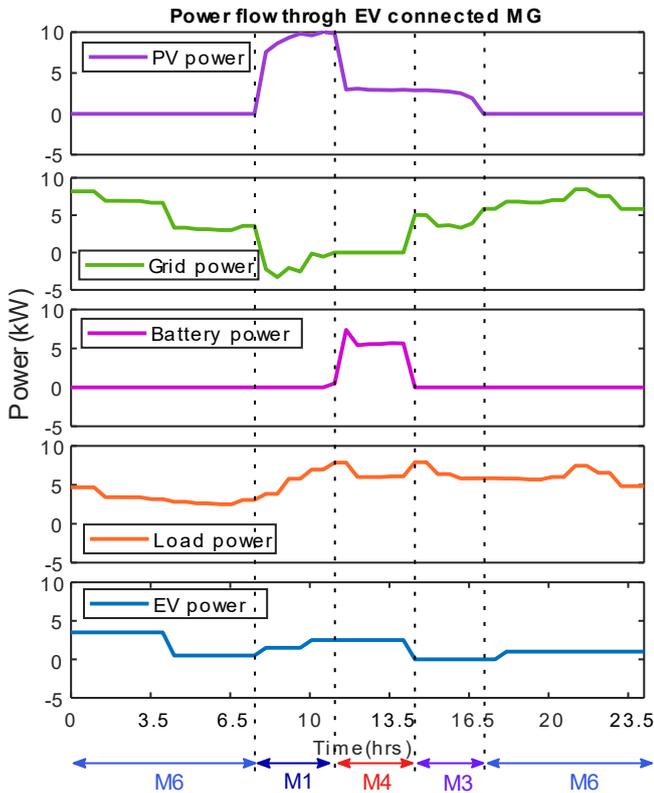

Figure 8. Power flow through the MG system with real time PV and load data.

I. CONCLUSION

This work proposed an EMS algorithm for a grid-connected PV-based MG system. The proposed EMS maintains the power flow to control battery charging/discharging and the use of additional PV generation. The algorithm takes decisions based on next-day solar irradiance forecasting, the day ahead loads demand and tariff profile. The focus of this EMS is to use either PV generation or lower tariff during to charge battery and higher tariff periods for battery discharging to avoid buying power from the grid. Also, this EMS takes decision to sell addition PV power generation during higher tariff period to avoid wastage of PV power, when neither load nor battery needs that much power generating by solar system. The validity of the proposed algorithm has tested using MATLAB/Simulink.